# Triclinic BiFeO₃: A room-temperature multiferroic phase with enhanced magnetism and resistivity

Md Sariful Sheikh [a, 1, †, *], Tushar Kanti Bhowmik [a, b, †, *], Alo Dutta [b], Sujoy Saha [c], Chhatra R. Joshi [d], T. P. Sinha [a]

[a] *Department of Physics, Bose Institute, 93/1, A.P. C. Road, Kolkata-700009, India*
[b] *Department of Condensed Matter Physics and Material Sciences, S.N. Bose National Centre for Basic Sciences, Block-JD, Sector-III, Salt Lake, Kolkata, 700106, India*
[c] *Department of Physics, Oakland University, Rochester, MI, 48309, USA*
[d] *Department of Physics and Astronomy, The University of Alabama, Tuscaloosa, AL 35487, USA*

## Abstract:

The magnetic and transport properties of BiFeO₃/La₂NiMnO₆ (BFO/LNMO) composite have been investigated both experimentally and theoretically. Unlike the normal rhombohedral (R3c) phase, BFO in the composites is crystallized in the triclinic phase (P1). Interestingly, the composites demonstrate a significant enhancement in the magnetization, magnetoelectric coupling and show higher resistivity than that of the regular BFO (R3c). As LNMO has its Curie temperature at 280 K, the room temperature and above room temperature magnetic contribution in the composites is expected to be from the triclinic BFO phase. Experimentally observed enhancement in magnetization is validated using classical Monte Carlo simulation and density functional theory (DFT) calculations. The calculations reveal higher magnetic moments in triclinic BFO as compared to the rhombohedral BFO. Overall, this study reveals triclinic BFO as the promising room temperature multiferroic phase which is helpful to optimize the multiferroicity of BFO and achieve wider applications in future.

[1] Present address: Department of Materials Science and Engineering, University of Wisconsin Madison, 1509 University Ave, Madison, WI 53706, United States
[†] Authors contributed equally to this work.
*Corresponding author, Email- physics.tushar@gmail.com, sarifulsekh@gmail.com

# I. Introduction:

ABO$_3$ type multiferroic perovskite oxide BiFeO$_3$ has attracted a great research attention owing to the rare coexistence of ferroelectric and anti-ferromagnetic ordering with high transition temperatures (ferroelectric Curie temperature, T$_c$ ~ 1100 K and Neel temperature, T$_N$ ~ 640 K) [1-3]. Since the initial demonstration of robust coupling between these two order parameters, the promising possibility of various fascinating applications in data storage systems, spintronics, microelectronics, and sensors has been explored [4-7]. BiFeO$_3$ shows highly distorted rhombohedral (R3c) perovskite structure at room temperature and G-type antiferromagnetic ordering, which originates from the super exchange interactions between the neighbouring half–filled d-orbital of Fe$^{3+}$ cation and occupied 2p-orbital of O$^{2-}$ anion [8]. Recently, exploiting these properties, low energy consuming non-volatile magnetoelectric spin-orbit device has been proposed [9]. However, the high leakage current due to low resistivity, low saturation magnetization, low remnant magnetization and weak magnetoelectric coupling in BiFeO$_3$ are the issues to be addressed before its realistic applications in the next generation logic and memory devices [1, 10-12]. The interplay among the crystal structure, charge, spin, and orbital degrees of freedom provides a fertile research ground for improving the multiferroic properties in BFO [7, 13, 14]. Modifying the crystal structure, spin ordering, and band gap are keys to acquiring better electrical and magnetic properties. Various approaches including, the elemental substitution at the A and/or B sites, high quality crystal formation, reducing the particle size, hetero-structure and solid solution formation have been turned out to be very effective to improve the multiferroic properties of BFO [8, 15-16]. Although promising, the multiferroicity and magnetoelectric coupling in either single phase BFO or composites/heterostructure are required to be further improved for its practical implementation, which demand more fundamental understanding of its structure-property correlation and its exploitation.

In this work, an experimental investigation combined with the theoretical calculations using the density functional theory (DFT) and Monte Carlo simulation (MCS) have been performed on the magnetic and electric properties of the $_{(1-x)}$BFO + $_{(x)}$La$_2$NiMnO$_6$ (LNMO) (x = 0.0, 0.1, 0.2 and 0.3) composites. The double perovskite oxide LNMO is another widely studied below room temperature ferromagnetic semiconductor (Curie temperature, T$_C$ ~ 280 K) with rich physical properties [17-18]. In our previous study we have observed that the large lattice parameters mismatch between BFO and LNMO lattice makes both the lattices highly strained and pushes the BFO lattice from being regular rhombohedral (R3c) to the lowest possible triclinic (P1) phase [19]. Previously, Chen et al. observed the formation of ferroelectric triclinic phase in the mixed phase regions of highly strained BiFeO$_3$ thin film and predicted large piezoelectric response in it [20]. Though the synthesis of multiferroic BFO based composites/heterostructures is a very effective approach to improve its multiferroic properties, most of the studies are on the regular rhombohedral, and strain-mediated monoclinic, orthorhombic, and tetragonal BFO systems [21-23]. But, the magnetic and electric properties of the triclinic BFO are yet to be fully explored. To get the fundamental insight into the electronic properties, we have also performed the DFT calculation on the triclinic BFO and compared it with that of the rhombohedral BFO.

The combined experimental and theoretical study demonstrates that triclinic BFO structure (P1) has the promising potential to overcome the low magnetism, weak magnetodielectric coupling and high leakage current of rhombohedral BiFeO$_3$. The modified structure of BFO (P1) possesses increased electrical resistivity, and an increased spin canting resulting in the significantly enhanced magnetisation as compared to the conventional R3c BFO.

The remaining part of this paper is organized as follows. In Sec. II, we have reported the experimental details. In Sec. III, we have represented the main results and its discussions on the findings. Our conclusions based on the obtained results are summarized in Sec. IV, while the sample preparation, Monte-Carlo simulation method, DFT calculation details and spontaneous polarisation calculations results are presented to the Appendixes A, B, C and D, respectively.

## II. Methodology:

The composites are synthesised using sol-gel method and a description of the materials synthesis and pellet preparation procedure are represented in Appendix A. The electrical measurements are performed using circular discs (8 mm diameter and ~ 1 mm thick) of the BFO/LNMO composites. The vibrating sample magnetometer (EverCool Quantum Design VSM magnetometer; Lakeshore) is used to study the magnetization of the as prepared pellets in the temperature range from 80 K to 400 K at zero field and an applied magnetic field of 5 kOe. The magnetic hysteresis properties are studied at 80 and 300 K. For the electrical measurements, the pellets were polished on both sides and silver electrodes were deposited using ultrapure silver paste (Ladd Research Industries. Inc). The capacitance ($C$) and conductance ($G$) were measured in the frequency range from 42 Hz to 1.6 MHz and in the temperature range 303 to 683 K using an LCR meter (3532-50, Hioki). The real part of the complex dielectric constant ($\varepsilon' = C/C_o$, where $C_o = \varepsilon_o A/t$, $\varepsilon_o$ is the permittivity of free space, $t$ is the pellet thickness, and $A$ is the surface area), imaginary part ($\varepsilon'' = Gx/\omega\varepsilon_o$, $\omega$ is the angular frequency) of the complex dielectric constant, dielectric loss factor (tan$\delta = \varepsilon''/\varepsilon'$) and Ac electrical conductivity ($\sigma = Gx$, $x = t/A$) were obtained from the frequency dependence of $C$ and $G$.

The electronic structure of BiFeO$_3$ has been investigated using the full potential linearized augmented plane wave as implemented in the WIEN2K code [24, 25]. The generalised gradient approximation (GGA) with the Hubbard parameter (U) method has been taken to study the spin polarised electronic band structure calculations. The threshold energy between valence and core states is fixed to -7 Ry for both triclinic and rhombohedral structures. The energy and charge cut-off are set to $10^{-4}$ Ry and $10^{-3}$ e for the self-consistent convergence in the scf cycles. The effective U value for the strong correlation between Fe-3d orbital electrons is set to 6 eV for both the cases. The temperature dependent magnetic properties of BFO/LNMO composite are analysed through the Monte-Carlo simulation method [26]. We have considered the anisotropic 3D-Ising model for the simulation and the Hamiltonian with nearest- neighbour (nn) is described as

$$H = -\sum_{<i,j>} J_{ij} s_i s_j - \Delta\sum_i s_i^2 - h\sum_i s_i,$$

Where $s_i$ and $s_j$ are the spins at lattice sites $i$ and $j$ respectively. $\Sigma_{<i,j>}$ is the summations made over spin pairs coupled through the nn interaction constant $J_{ij}$ and the magnetocrystalline anisotropy energy constant ($\Delta$) and $h$ is the external magnetic field applied along z- axis. The detailed Monte-Carlo simulation and DFT calculations methods are described in the Appendix B and C, respectively.

## III. Results and discussion:

## A. Structural property

The detailed structural characteristics are reported in our previous work [19], and a summary of the crystal structures and lattice parameters of the BFO phases are represented in Table 1 and Figure 1. The pure BiFeO$_3$ (x = 0.0) is crystallised in the usual rhombohedral (space group R3c; a = b = c = 5.643 Å, α = β =90°, γ = 120°) phase as shown in Figure 1(a). But in the BFO/LNMO composite BFO is crystallised in the triclinic (space group P1; a = 3.931 Å, b = 3.936 Å, c = 3.956 Å, α = 90.32°, β = 90.24°, γ = 89.98°) phase as shown in Figure 1(b). Figure 1(c) shows a qualitative comparison of the unit cell shape of the BFO structures. The double perovskite oxide LNMO is crystallized in the ordered monoclinic structure which is further supported by the magnetic study as discussed in Section IIIC. We also observed similar strain mediated structural transition in multiferroic KNbO₃/LNMO composite, where the large lattice strain due to the lattice parameter mismatch transformed orthorhombic KNbO₃ to its cubic perovskite structure [27].

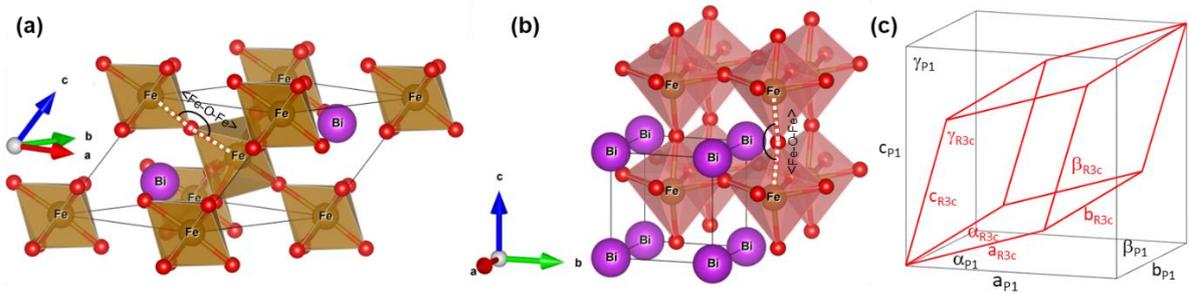

**Figure 1:** Unit cell representation of (a) Rhombohedral BFO and (b) triclinic BFO. (c) Comparison of unit cells of R3c (red) and P1 (black) BFO structure.

**Table 1**: Lattice parameters of the BiFeO₃ phases in the (1-x)BiFeO₃ + (x)La₂NiMnO₆ composites [19].

| Lattice parameters | x = 0.0 BFO (R3c) | x = 0.1 BFO (P1) | x = 0.2 BFO (P1) | x = 0.3 BFO (P1) |
|---|---|---|---|---|
| a (Å) | 5.643 | 3.931 | 3.907 | 3.881 |
| b (Å) | 5.643 | 3.936 | 3.952 | 3.939 |
| c (Å) | 5.643 | 3.956 | 3.928 | 3.914 |
| Cell volume (Å³) | 125.18 | 61.34 | 60.85 | 59.86 |
| Average Fe-O bond length (Å) | 2.01 | 2.039 | 2.0496 | 2.043 |
| Average FeO₆ octahedra volume (Å³) | 10.845 | 10.224 | 10.142 | 9.987 |

## B. Electrical conductivity

Figure 2(a) shows the logarithmic angular frequency, log$(\omega)$ dependence of the AC conductivity, $\sigma(\omega)$ which exhibits the typical semiconductor behaviour with a plateau region in the low frequency region. The conductivity value in the low frequency plateau is nearly equal to the DC conductivity value ($\sigma_{dc}$) and the conduction mechanism is also similar to its DC conduction mechanism i.e., long range translational hopping of the thermally generated charge carriers from one localised site to another [28]. With the increase in frequency the probability of successful long range hopping decreases and the probability of the short-range hopping increases. As the frequency exceeds the critical hopping frequency ($\omega_H$), the ratio of the long range to short range hopping starts to decrease which results in a dispersive conductivity in the higher frequency region. The $\sigma(\omega)$ can be fitted and explained by the following Jonscher power law [29, 30],

$$\sigma(\omega) = \sigma_{dc} \left[ 1 + \left( \frac{\omega}{\omega_H} \right)^n \right]$$

where, $n$ is a dimensionless parameter varying from 0 to 1. The fitted values of $\sigma_{dc}$, $\omega_H$ and $n$ are shown in Table 2. The value of the critical frequency $\omega_H$ increases as the LNMO loading increases in the system as compared to the pure rhombohedral BFO. The conduction in BFO is governed by the Fe cation and oxygen vacancy [31]. The oxygen vacancy formation during the synthesis process causes reduction of a fraction of $Fe^{3+}$ cations and the formation of $Fe^{2+}$. The $Fe^{2+}$ cations act as the trap centre within the BFO lattice and considered responsible for the large leakage current. The interaction between thermally released electron and phonon in a polar medium like, BFO, result in the formation of localised small polaron within BFO lattice [32, 33]. The slow motion of the small polarons between the nearest neighbouring sites of BFO causes lattice distortion considered as polaron hopping. With increment in frequency the occupancy of the trap centre is reduced, and the density of the localised electrons increases making them available for the conduction. Hence the conductivity of the semiconductor system increases with the frequency as observed in Figure 2(a).

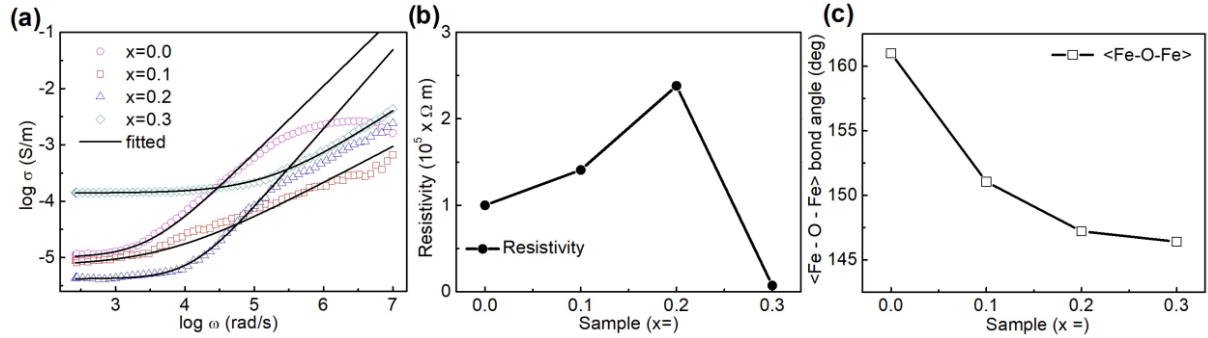

**Figure 2: (a)** Logarithmic angular frequency dependence of ac conductivity. Solid lines indicate the power law fitted data. (b) Room temperature dc-conductivity, and (c) <Fe – O – Fe> bond angle plot as a function of LNMO molar concentration.

The resistivity values ($1/\sigma_{dc}$) obtained from the $\sigma_{dc}$ values are plotted in Figure 2(b), which shows that with the LNMO loading the resistivity of the samples increases up to x = 0.2. This resistivity change may be attributed to the structural transformation from rhombohedral (R3c) BFO to triclinic (P1) BFO. The electronic structure and electrical properties in ABO$_3$ perovskites are determined by the hybridisation strength between Fe and O atoms in the FeO$_6$ octahedra. Figure 2(c) shows the <Fe – O – Fe> bond angle, which play a crucial role in the magnetic and electric properties of BFO, significantly reduces from rhombohedral to the triclinic BFO in the composites. The reduced <Fe – O – Fe> bond angle decreases the hybridisation strength between Fe-3d and O-2p orbitals in triclinic BFO which is responsible for the increased resistivity in the composites. On the other hand, the excess loading of the relatively narrow band gap LNMO forms interconnected electrical networks which results in the reduced resistivity in x = 0.3.

**Table 2:** Jonscher power law fitting parameters.

| Sample | $\sigma_{dc}$ x 10$^4$ (S/cm) | $\omega_H$ (rad/s) | n |
| --- | --- | --- | --- |
| x = 0.0 | 10.0 | 2900 | 0.9 |
| x = 0.1 | 14.1 | 5500 | 0.91 |
| x = 0.2 | 23.8 | 12500 | 0.3 |
| x = 0.3 | 71.4 | 165000 | 0.8 |

## C. Magnetic property

The low temperature dependence of the magnetization plots (in the temperature range 80-400 K) (Figure 3(a)) of the samples at 5 kG shows a large increment in the magnetization in the composites as compared to the pure rhombohedral BFO. A sudden rise in the magnetisation at temperature ~280 K has been observed in the composites as the temperature decreases, which is related to the ferromagnetic transition of LNMO. In general, the LNMO lattice offers both the long-range ordering of Ni and Mn cations and forms the ordered monoclinic (P2$_1$/n) or rhombohedral (R$\overline{3}$m) structures; and disordered orthorhombic (Pbnm) phase [34, 35]. Ordered LNMO phases have the Ni$^{2+}$/Mn$^{4+}$ oxidation states and the Ni$^{2+}$–O–Mn$^{4+}$ super exchange interaction generates ferromagnetism with the Curie temperature (T$_{CM}$ ~ 280 K) just below the room temperature. Whereas disordered LNMO phase has trivalent Ni$^{3+}$/Mn$^{3+}$ oxidation states and the Ni$^{3+}$–O–Mn$^{3+}$ vibronic super exchange interaction incorporates ferromagnetism with a lower T$_{CM}$ (~ 150 K) [36]. Hence, the Curie temperature (T$_C$~280 K, as shown in Figure 3(a)) suggests the existence of ordered LNMO phase in the composites and the magnetic contribution above 280 K is completely attributed to the other source of origin which is in this case triclinic BFO phase. The plots of the magnetisation (M) as a function of magnetic field (H) at temperatures 80 and 300 K are shown in Figure 3(b) and Figure 3(c), respectively. Pure BFO presents the simple linear M(H) dependence with very small remanent magnetization (M$_r$) and saturation magnetization. The summary of the magnetic parameters at 300 and 80 K are tabulated in Table 3 and 4. It is noteworthy to mention here that a large increment in the magnetisation has been observed in the composites. The remanent magnetization for the samples with x = 0.0, 0.1, 0.2 and 0.3 are 0.0003, 0.304, 0.425, 0.332 emu/gm, respectively. The room temperature remanent magnetization in the samples with x = 0.1, 0.2 and 0.3 are almost increased by 1000,1400 and 1100 times, respectively, as compared to the pure BFO. The room temperature magnetization at 14 kG for the samples x = 0.0, 0.1, 0.2 and 0.3 are 0.09, 0.92, 1.55, 1.50 emu/gm, respectively. The magnetization for the samples x = 0.1, 0.2 and 0.3 are increased by 10.2, 17.2 and 16.6 times, respectively. The comparison of the room temperature remanent magnetization of (0.8BFO + 0.2LNMO) with some other reported BFO based bulk samples as tabulated in Table 5, suggests that this triclinic BFO has much higher magnetization as compared to other phases of BFO. However, the lowering of BFO concentration may be the reason for decreasing magnetisation in the sample with x = 0.3. The overall increment in magnetisation in the composites may be because of the triclinic crystal structure of BFO. The magnetisation in multiferroic BFO depends on the antiferromagnetic sublattice angularity, with the Dzyaloshinski-Moriya interaction (DM) playing the key role. The DM interaction leads to the spin canting resulting in the appearance of a rather small total FM moment in an AFM structure, thus enhancing the magnetism. Here the decreased <Fe – O – Fe> bond angle in the triclinic BFO increases the canting of spins and, hence, improves the magnetization as compared to the R3c BFO [37-39].

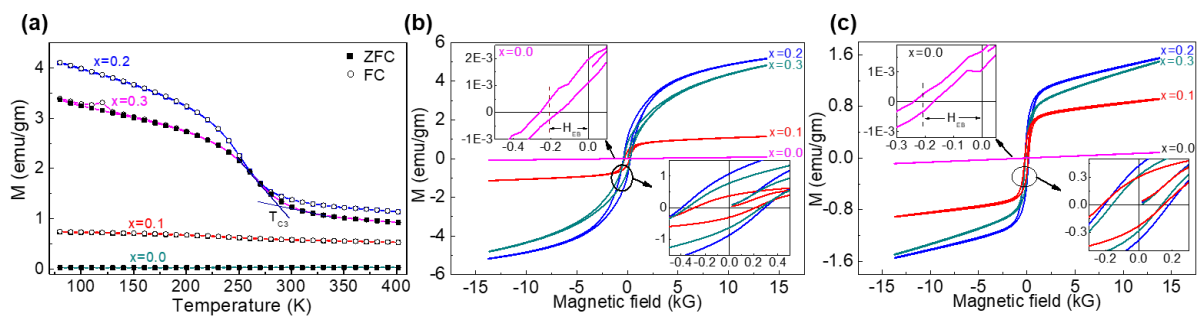

**Figure 3:** (a) Temperature dependent magnetisation plots in both zero-field cooling and field cooling condition (5 kG), Experimental MH loop at (b) 80 K and (c) 300 K. The figures in the inset has the same unit as the main figure.

**Table 3:** Magnetic parameters derived from the experimental and Monte-Carlo (M-C) simulated M-H hysteresis loop at 300 K.

| Sample | $M_r$ (emu/gm) | | $M_s$ (emu/gm) | | $H_{cr}$ (emu/gm) | |
|--------|-------|-------|-------|-------|-------|-------|
| | Expt. | M-C | Expt. | M-C | Expt. | M-C |
| x = 0.0 | 0.0003 | 0.000024 | 0.088 | 0.004 | 0.035 | 0.007 |
| x = 0.1 | 0.304 | 0.020 | 0.922 | 0.913 | 0.188 | 0.137 |
| x = 0.2 | 0.425 | 0.480 | 1.551 | 1.167 | 0.195 | 0.149 |
| x = 0.3 | 0.332 | 0.32 | 1.498 | 1.409 | 0.149 | 0.091 |

**Table 4:** Magnetic parameters derived from the experimental and Monte-Carlo (M-C) simulated M-H hysteresis loop at 80 K.

| Sample | $M_r$ (emu/gm) | | $M_s$ (emu/gm) | | $H_{cr}$ (emu/gm) | |
|--------|-------|-------|-------|-------|-------|-------|
| | Expt. | M-C | Expt. | M-C | Expt. | M-C |
| x = 0.0 | 0.0005 | 0.0008 | 0.087 | 0.092 | 0.051 | 0.096 |
| x = 0.1 | 0.383 | 0.404 | 1.149 | 0.61 | 0.252 | 0.486 |
| x = 0.2 | 1.076 | 4.029 | 5.17 | 4.04 | 0.347 | 2.486 |
| x = 0.3 | 0.766 | 3.406 | 4.822 | 3.417 | 0.302 | 1.452 |

**Table 5:** Comparison of the magnetic moment of sample with x = 0.2 with some recent bulk BFO based magnetic materials.

| Material | $M_r$ (emu/gm) | Year | temperature |
|----------|-------|------|-------------|
| 0.8BFO + 0.2LNMO | 0.425 | This work | 300 K |
| 0.7BFO + 0.3TbMnO$_3$ [40] | ~ 0.15 | 2019 | 275 K |
| $Bi_{0.95}Gd_{0.05}FeO_3$ [41] | 0.218 | 2019 | RT |
| $Bi_{1-x}Ca_xFeO_3$ (x=0.1) [42] | 0.096 | 2019 | RT |
| $0.8(Bi_{0.9}La_{0.1}FeO_3)–0.2(KBr)$ [43] | ~ 0.25 | 2019 | 300 K |
| $BiFe_{1-x}Se_xO_3$ x=0.25 [44] | 0.034 | 2019 | 300 K |
| $Bi_{0.92}La_{0.08}Fe_{1-x}Se_xO_3$, x=0 [44] | 0.1334 | 2019 | 300 K |
| $Bi_{0.5}La_{0.5}Fe_{0.5}Mn_{0.5}O_3$ [45] | 0.7 (Much lower at RT) | 2017 | 2 K |

## D. Monte-Carlo simulations of magnetic properties

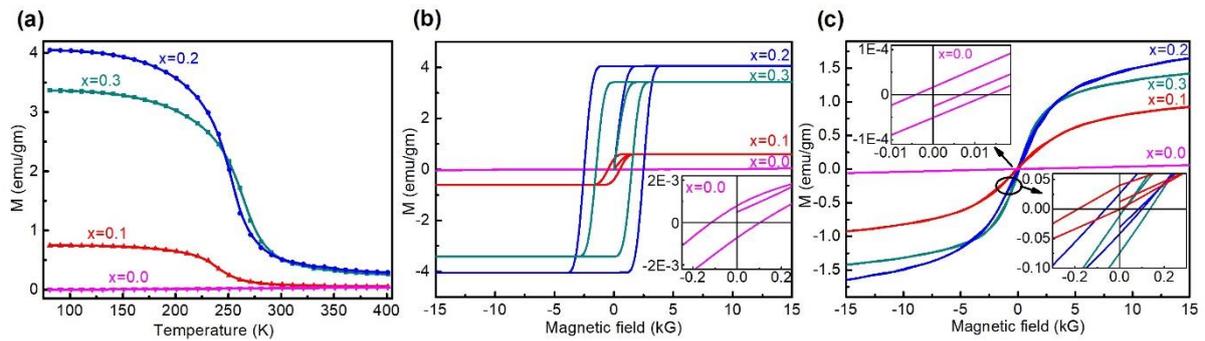

**Figure 4:** (a) Monte-Carlo simulation of magnetisation versus temperature. Monte-Carlo simulated MH loop at (b) 80 K and (c) 300 K. The Figures in the inset has the same unit as the main Figure.

In our study, we have also performed the Monte Carlo simulations using the Metropolis algorithm to investigate the magnetic properties of the BFO/LNMO solid solutions and the detailed calculation formalism are described in the Appendix B [46-48]. As shown in Figure 4(a), the simulated temperature-dependent magnetization plots of the systems follow the same pattern as the experimental data. This indicates that our simulations accurately capture the magnetic behaviour of the systems under investigation. Figure 4(b) shows that the M-H loops at 80 K manifest the FM nature of the composites compared to the AFM R3c BFO. Therefore, the FM nature of the composites may be attributed to the FM nature of LNMO and triclinic BFO. Furthermore, we have also investigated the effect of external magnetic fields on the room temperature hysteresis loops of the BFO/LNMO composite systems. The simulated MH loops above the Curie temperature of LNMO, i.e., at 300 K, as shown in Figure 4(c) also suggest the ferromagnetic nature of the solid solutions which is consistent with the FM nature of P1 BFO. Moreover, the simulations showed a considerable increase in magnetization compared to pure BFO, which is qualitatively consistent with the experimental data as represented in Table 3 and 4. These findings highlight the success of our Monte Carlo simulations in accurately reproducing the temperature-dependent magnetization behaviour and the influence of external magnetic fields on the BFO/LNMO composite systems.

### E. Dielectric properties

We also studied the temperature dependent dielectric constant ($\epsilon'$) of samples x=0 and 0.2 to qualitatively check the magnetoelectric coupling. The appearance of a dielectric anomaly near the magnetic transition temperature of a multiferroic material indicates the presence of the multiferroic magnetoelectric coupling [49]. If the $\epsilon'$ shows a minute anomaly at the magnetic transition temperature the dielectric property of this material is generally less sensitive to the magnetic field. As the ferroelectric and magnetic ordering in $BiFeO_3$ are associated with different ions Bi and Fe, respectively, the multiferroic coupling in $BiFeO_3$ is relatively weak [50]. The temperature dependent dielectric constant plots of the rhombohedral BFO and the composite with x = 0.2 as shown in Figure 5(a) reveal the dielectric anomaly near the magnetic transition temperature suggesting the presence of the magnetoelectric coupling below 640 K in these materials. Interestingly, the composite with x = 0.2 shows significantly enhanced dielectric anomaly near the magnetic transition temperature at around 640 K suggesting much stronger magnetoelectric coupling in it as compared to the R3c BFO. But more research on the spin structure-property corelation of triclinic BFO is required to understand this behaviour. We have also calculated the dielectric loss tangent (tanδ), which quantitatively denotes the dissipation of electrical and electromagnetic energy (e.g., heat) due to different physical process such as electrical charge transport, dielectric relaxation, resonant transition, and non-linear dielectric effects in a dielectric medium [51]. The dielectric loss tangent also shows the anomaly near the magnetic transition temperature which is consistent with the dielectric constant plot. The dielectric loss tangent, tanδ of sample with x = 0 and x = 0.2 as shown in Figure 5(b) demonstrates nearly similar energy dissipation up to 400 K, and much lower dielectric loss in the composite with x = 0.2 as compared to x = 0 above 400 K. Overall, these experimental studies reveal the significantly higher magnetization and stronger magnetoelectric coupling in the triclinic BFO based composite, motivating us to further explore this triclinic BFO structure using DFT calculations formalism.

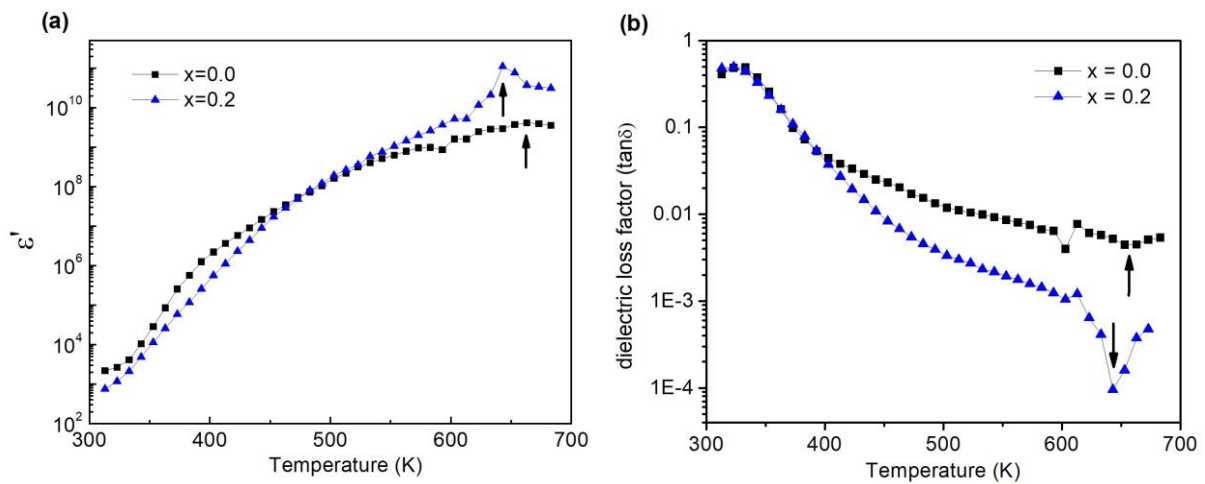

**Figure 5:** (a) Temperature dependence of the real part of the dielectric constant (ε'), and (d) dielectric loss factor at frequency 4.8 MHz.

## F. Density functional theory calculations

We have studied the band-structure, density of states (DOS), magnetic moments and spontaneous polarisation in both R3c and P1 BFO, which are described below. The spin polarised band structure of the rhombohedral BFO is shown in Figure 6(a, b), and the band structure of triclinic BFO is presented in Figure 6(c, d). The DFT calculated lowest band gap for the R3c BFO phase for both the spin-up and spin-down conditions is 1.92 eV, consistent with the experimental direct band gap of bulk rhombohedral BFO (~ 2 eV) [52]. On the other hand, the triclinic phase has band gaps of 2.25 eV and 2.22 eV, as observed in the spin-up and spin-down conditions, respectively. The lowest band gap in P1 BFO is in the down spin channel (Figure 6(d)) which is higher than the R3c BFO. In the non-optical applications of multiferroic materials electron can jump from valence band to conduction band via thermal excitation i.e., by phonon absorption. As, phonon have zero spin, the electron generally conserves its spin during thermal excitation from the valence band to the conduction band. Hence the lowest energy gap for P1 BFO for the non-spin flip thermal excitation is 2.2 eV, higher than R3c BFO (1.92 eV). Hence, due to the wider energy gap the triclinic BFO is expected to possess higher resistivity than the rhombohedral BFO, which is consistent with the experimental DC conductivity study. This observed higher resistance may help to reduce the leakage current of BFO, which has been a main shortcoming of BFO for its device use. Interestingly, the spin flip optical bandgap of P1 BFO is 1.56 eV, which is lower than the R3c BFO (1.92 eV). Besides multiferroism, BFO has shown its promising potential for various photo sensing applications including solar cell, photoelectrochemical water splitting, etc., in recent years [28, 53,54]. However, its efficiency in the photo sensing devices is limited due to the wide optical band gap and related poor visible light absorbance of R3c BFO. Hence, P1 BFO with its relatively narrower optical band gap, which is much closer to the Shockley-Queisser band gap (~1.34 eV) [55], could significantly enhance the performance of BFO in photo sensing devices.

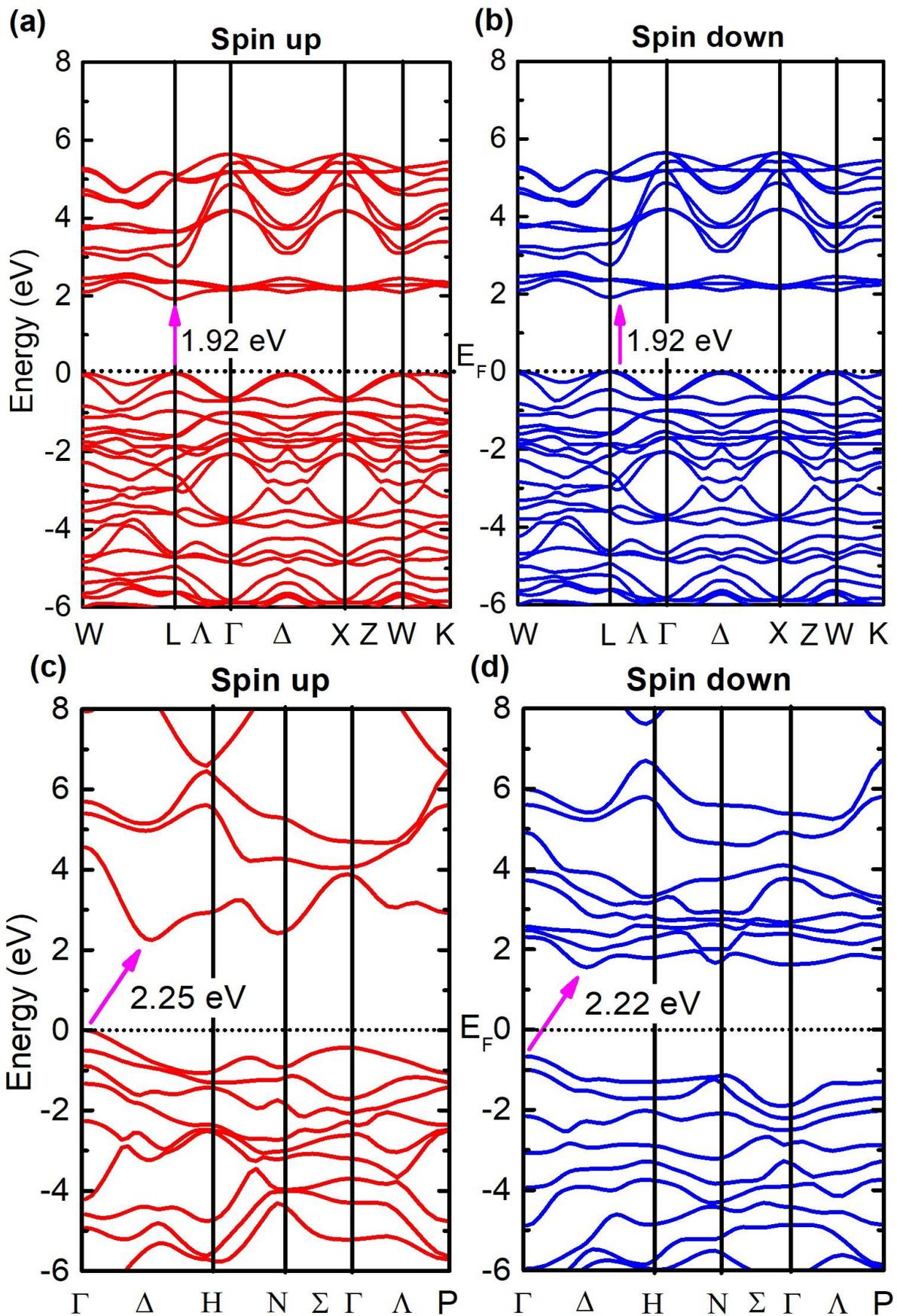

**Figure 6:** Band structure of rhombohedral BFO for (a) spin up and (b) spin down configuration. Band structure of triclinic BFO for (c) spin up and (d) spin down configuration.

Figure 7(a) and (b) display the spin-polarized DOS for the R3c and P1 structures of BFO, respectively. Upon analysing the partial DOS, it becomes evident that the density in the valence band maximum (VBM) and conduction band minimum (CBM) is primarily dominated by the Fe-3d and O-2p states, with a minor contribution from the Bi-6p state in both the R3c and P1 structures of BFO. However, it is noteworthy that the overlap between the O-2p and Fe-3d states is more pronounced in the triclinic BFO structure compared to the rhombohedral BFO structure. This observation suggests a stronger covalent nature of the Fe-O bonds in the triclinic BFO configuration. The spin-polarised symmetric DOS of R3c BFO reveals its conventional AFM phase. On the other hand, the spin-polarised asymmetric DOS of P1 BFO again confirms its FM nature, consistent with the experimental results. In terms of magnetic moments, the calculated values for the Bi, Fe, and O atoms in the rhombohedral BFO structure are found to be 0.049 µB (±0.027 emu/gm), 3.56 µB (±355.85 emu/gm), and 0.33 µB (±16.75 emu/gm), respectively. It is worth mentioning that the calculated magnetic moment of the $Fe^{3+}$ ion is consistent with both previous experimental and theoretical results. A low-temperature neutron-diffraction study conducted on the R3c BFO structure reported a measured magnetic moment for the $Fe^{3+}$ ion of 3.75 µB (374.85 emu/gm) [56], which aligns closely with the calculated value reported in this work. Additionally, a previous DFT study also demonstrated that the magnetic moment of the $Fe^{3+}$ ion in the R3c BFO structure is approximately 3.65 µB (364.85 emu/gm) [57].

**Table 6:** DFT calculated magnetic moments of R3c and P1 BFO.

| Ions | R3c (emu/gm) | P1 (emu/gm) |
|---|---|---|
| $Bi^{3+}$ | ± 0.03 | 0.37 |
| $Fe^{3+}$ | ± 355.85 | 412.83 |
| $O^{2-}$ | ± 16.75 | 41.87 |
| Interstitial | 0.00 | 29.77 |

| Cell | 0.00 | 356.88 |
| --- | --- | --- |

In the triclinic P1 structure of BFO, the spin magnetic moments for the Bi, Fe, and O atoms are determined to be 0.014 µB (0.373 emu/gm), 3.97 µB (412.83 emu/gm), and 0.19 µB (41.87 emu/gm), respectively. A comparison of the DFT calculated magnetic moments of both R3c and P1 BFO are represented in Table 6. These values indicate an increased magnetic moment in the triclinic BFO structure compared to rhombohedral structure. The theoretical spin-only magnetic moment for the $Fe^{3+}$ ion is calculated to be 5.91 µB (590.76 emu/gm), based on the formula $\sqrt{n(n+2)}$, where n represents the number of unpaired electrons. However, the calculated magnetic moment from DFT calculations is found to be lower than the theoretical value. This deviation can be attributed to the hybridization between the Fe-3d and O-2p orbitals, where the five 3d electrons of $Fe^{3+}$ do not localize on the atomic orbitals but rather form Wannier orbitals, where the presence of spin nonpolarized O 2p states makes a notable contribution which affects the magnetic moment [58].

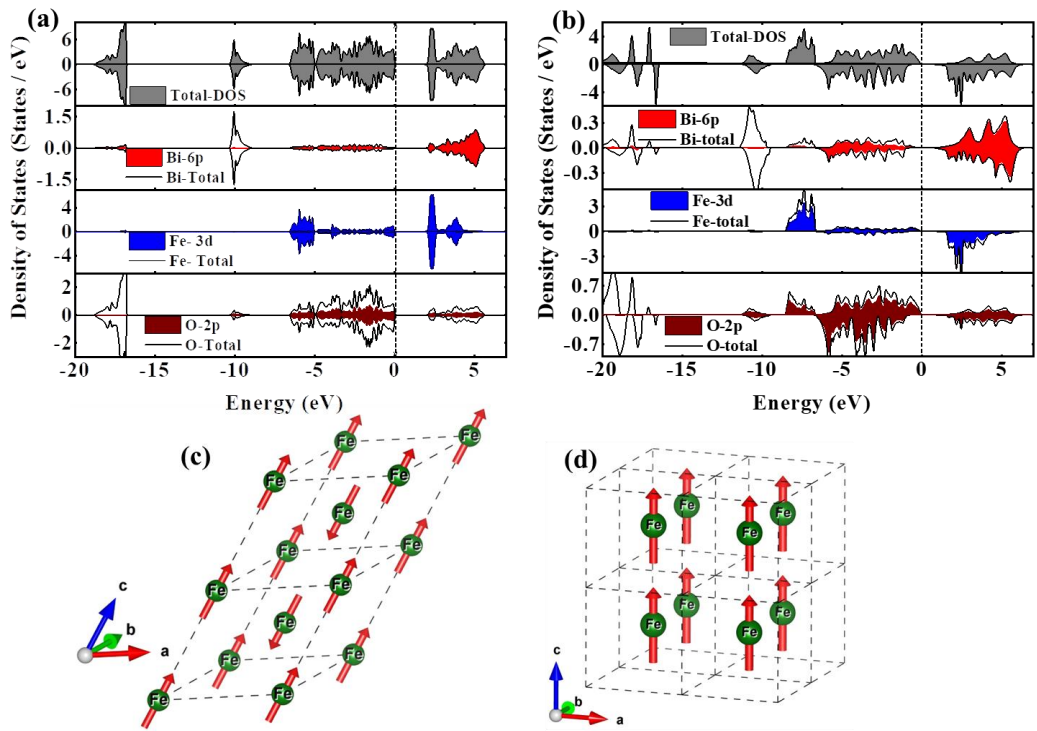

**Figure 7:** Spin polarised DOS and PDOS of (a) R3c, and (b) P1 BFO. Ground state magnetic structure of (c) R3c and (d) P1 BiFeO$_3$ derived from DFT calculation.

To visualize the magnetic spin structure, ground state magnetic properties of R3C and P1 structures were determined using DFT calculations as depicted in Figure 7(c) and (d). These figures provide a representation of the spatial distribution and orientation of the magnetic spins in the materials, aiding in the understanding of its magnetic behaviour. These findings further provide valuable insights into the spin-polarized DOS and magnetic moments of BFO in both the crystal structures. Overall, the analysis of the magnetic moments and spin structures in the triclinic BFO structure shed light on its magnetic properties and effects of orbital hybridization on magnetic moment.

In summary, the strain mediated triclinic (P1) BFO phase demonstrated enhanced magnetism along with increased resistance and magnetoelectric coupling as compared to the regular R3c BFO phase. Hence, the next focus should be on the strain-mediated synthesis of 2D triclinic BFO on a suitable lattice matched substrate and its exploration. The 2D triclinic BFO based heterostructure is expected to demonstrate improved resistivity, high saturation magnetisation, remanence magnetisation and robust magnetoelectric coupling which is essential for the practical implementation of room temperature multiferroic technology.

## IV. Conclusions:

The studies on the electronic and magnetic properties of the composite BFO/LNMO have been performed, and the combined experimental and theoretical analysis demonstrate the remarkable improvement of the magnetization and multiferroic magnetoelectric coupling in the composites as compared to the rhombohedral BFO. The enhanced room temperature magnetisation and resistivity in the composites are attributed to the triclinic BFO phase. Interestingly, triclinic BFO has decreased <Fe - O - Fe> bond angle and an increased spin canting those results in the significantly increased resistivity and enhanced magnetisation, respectively, as compared to the regular BFO (R3c) structure. Overall, this study introduces triclinic BFO (P1) phase as a promising pathway to realize room temperature multiferroic material and opens a new avenue to overcome the limiting factors of the current multiferroic technology.

**Acknowledgements:**

Md S. Sheikh and T. K. Bhowmik would like to thank the Department of Science and Technology (DST), Government of India for providing the INSPIRE Fellowship (Sanction No: IF150220 and IF160418, respectively).

**Appendix A: (1-x)BFO + (x)LNMO  (x = 0.0, 0.1, 0.2, 0.3) synthesis and pellet preparation**

The multiferroic composite (1-x)(BFO) + x(LNMO) (with x = 0.0, 0.1, 0.2 and 0.3) powders were prepared using sol-gel method [19]. First, the reagent grade metal nitrates, $Bi(NO_3)_3.5H_2O$, $Fe(NO_3)_3.6H_2O$, $La(NO_3)_3.3H_2O$, $Ni(NO_3)_2.4H_2O$ and $Mn(NO_3)_2.6H_2O$ were dissolved in 2-methoxy ethanol solvent. 3% of extra $Bi(NO_3)_3.5H_2O$ was taken to avoid the Bi evaporation loss during the heating process. A solution of tartaric acid in 2-methoxy ethanol (in 1:1 molar ratio with the metal nitrates) was added drop wise in the first solution and mixed well. The well mixed solutions were dried at 393 K to get the black precursor powder. Finally, the precursor powders were heated in ambient furnace in two steps heating method. In the first step, the samples were heated at 723 K for 2 hours to ensure the complete phase formation of BFO. In the second step, the furnace temperature was gradually increased to 923 K from 723 K and heated for 5 hours to ensure the complete double perovskite phase formation of LNMO. Finally, the furnace was cooled down to the room temperature at a cooling rate of 1 K/min. The synthesised powders were palletised into circular discs of average diameter 100 mm and thickness 2 mm using the polyvinyl alcohol as the binder and sintered at 973 K for 3 hours for the magnetic and dielectric measurements.

**Appendix B: Model and Monte-Carlo Simulation details**

The temperature dependent magnetic properties of BFO/LNMO composite are analysed through the Monte-Carlo simulation method. For the triclinic BFO structure, we have considered the only one magnetic ion $Fe^{3+}$ ion at corner edges [26]. We have considered the anisotropic 3D-Ising model for the simulation whereas any other models like Heisenberg model does not satisfy the experimental data so accurately. So, the anisotropic 3D Ising model Hamiltonian with nearest- neighbour (nn) is described as

$$H = -\sum_{<i,j>} J_{ij} s_i s_j - \Delta \sum_i s_i^2 - h \sum_i s_i,$$

Where $s_i$ and $s_j$ are the spins at lattice sites $i$ and $j$ respectively. $\Sigma_{<i,j>}$ is the summations made over spin pairs coupled through the nn interaction constant $J_{ij}$ and the magnetocrystalline anisotropy energy constant ($\Delta$) -24.8 $\mu$eV/Fe is taken from previous study [59] and $h$ is the external magnetic field applied along z- axis. For the complex perovskite [(1-x)(BFO) + x(LNMO)] (x = 0.1, 0.2, 0.3), we have taken two sublattices; one is for $BiFeO_3$ and another for $La_2NiMnO_6$. The whole complex perovskite consists of the three magnetic ions, $Fe^{3+}$, $Ni^{2+}$ and $Mn^{4+}$. So, there are six spin spin interactions present in the system and the interaction constants are $J_{Fe-Fe}$, $J_{Ni-Ni}$, $J_{Mn-Mn}$, $J_{Ni-Mn}$, $J_{Fe-Ni}$, and $J_{Fe-Mn}$. But, the interaction between Fe-Ni and Fe-Mn are very small due to the small concentration of LNMO and large atomic distance between these atoms. So, the nearest neighbour interaction constants are $J_{(Fe-Fe)}$, $J_{(Ni-Mn)}$, $J_{(Ni-Ni)}$, and $J_{(Mn-Mn)}$. BFO has antiferromagnetic ordering temperature ($T_N$) around 640 K [60]. The LNMO has a paramagnetic to ferromagnetic transition around 280 K upon cooling. So, the interactions between Ni-Ni, Mn-Mn, and Ni-Mn are ferromagnetic. Masrour et. al. has determined the interaction constants for LNMO from the mean-field theoretical calculations [61]. We have taken the spin magnitude of $Fe^{3+}$, $Ni^{2+}$ and $Mn^{4+}$ are 5/2, 1 and 3/2 respectively. For pure BFO, we have determined the interaction constants from the mean field approximation. The formula is given by,

$$T_N = \frac{2}{3K_B} ZS(S+1)J_{Fe-Fe},$$

Where Z is the coordination number and S is the spin value of all magnetic ions. $T_N$ denotes the experimental value of the transition temperature and $J_{Fe-Fe}$ is the interaction constant and $K_B$ is the Boltzmann constant. For the pure R3c BFO the reported transition temperature is 640 K. The calculated value of $J_{Fe-Fe}$ in the case of the R3c BFO turns out to be 18.28 K. For the composite materials, we have calculated the interaction constant from the mean-field approximation formula, $T_C = \frac{2}{3K_B} ZS(S+1)J$. The transition temperature, $T_C$ in the composites were determined from their temperature dependent magnetisation data as shown in Figure 3(a). For x = 0.1, we have found the experimental $T_{C1} = 276$ K and the calculated value of the coupling constants are $J_{Fe-Fe} = 7.8$ K, $J_{Ni-Ni} = 34.5$ K, $J_{Mn-Mn} = 18.4$ K, $J_{Ni-Mn} = 26.1$ K. For x = 0.2 ($T_{C2} = 280$ K), we have found the value of $J_{Fe-Fe} = 8$ K, $J_{Ni-Ni} = 35.01$ K, $J_{Mn-Mn} = 18.7$ K, $J_{Ni-Mn} = 26.5$ K. In case of x = 0.3, the experimental $T_{C3} = 286$ K and we have calculated the value of $J_{Fe-Fe} = 8.2$ K, $J_{Ni-Ni} = 35.75$ K, $J_{Mn-Mn} = 19.07$ K, $J_{Ni-Mn} = 27.08$ K.

The magnetic properties of the BFO and BFO/LNMO composites are analysed using MCS under the Metropolis algorithm using the above-described Hamiltonian with cyclic boundary conditions on the lattice [62, 63]. We have performed this MCS on the sample of dimension N = L × L × L, where L is the number of unit cells in all three directions. We have taken N = 32 × 32 × 32 number of lattices for these simulations. The single-spin flip mechanism is used for all lattice sites to minimize the internal energy and the flips of these spin values are accepted or rejected according to the Boltzmann statistics [63]. At each temperature for every spin configuration, a number of $10^5$ Monte Carlo steps have been performed to equilibrating the lattice and next $10^6$ steps for thermal average of the total magnetisation, $M = \frac{1}{N} < \sum_i s_i >$. Where the sum is performed over all spin values of $Fe^{3+}$, $Ni^{2+}$ and $Mn^{4+}$ and $<...>$ indicates the statistical time average.

**Appendix C: DFT Calculation method**

The electronic structure of $BiFeO_3$ has been investigated using the full potential linearized augmented plane wave as implemented in the WIEN2K code [23, 24]. The generalised gradient approximation (GGA) with the Hubbard parameter (U) method has been taken to study the spin polarised electronic band structure calculations. The threshold energy between valence and core states is fixed to -7 Ry for both triclinic and rhombohedral structures. The maximum angular momentum, $l_{max}$ for the wave function expansion inside the atomic spheres is set at 10 and the plane wave cut off $R_{MT} \times K_{Max}$ is taken to 7 Ry with 1000 k points mesh integration over the first Brillouin zone to solve the Kohn- Sham equations. Here, the $R_{MT}$ and $K_{Max}$ stand for the average Muffin-Tin radii of the ionic sphere and the wave function cut-off. The Muffin-Tin radii of BFO under triclinic P1 space group are 2.16 Å, 1.69 Å and 1.46 Å for Bi, Fe and O atoms, respectively. Whereas for rhombohedral R3c they are 2.5 Å, 1.9 Å and 1.63 Å, respectively. The energy and charge cut-off are set to $10^{-4}$ Ry and $10^{-3}$ e for the self-consistent convergence in the scf cycles. The effective U value for the strong correlation between Fe-3d orbital electrons is set to 6 eV for both the cases.

We have employed the Quantum Espresso package [64] to perform calculations concerning the spontaneous polarization. The projector augmented-wave (PAW) method was utilized, along with the GGA based on the Perdew-Burke-Ernzerhof (PBE) functional, to account for exchange-correlation effects. A plane-wave energy cutoff of 75 Ry was set for the SCF calculation. For geometry optimization and electronic structure calculations, a $7 \times 7 \times 7$ Monkhorst-Pack k-point mesh was employed. Addressing the strong Coulomb repulsion (U) occurring between the localized d-states of Fe, we incorporated the GGA+Hubbard potential (U) approach to appropriately describe correlation effects within transition metal oxides. In this study, a Hubbard potential U value of 6 eV was applied to Fe after careful consideration.

**Table 7:** Lattice parameter used for the DFT calculations (before and after energy minimization)

| Symmetry | Condition | a (Å) | b (Å) | c (Å) | α (deg) | β (deg) | γ (deg) |
|----------|-----------|-------|-------|-------|---------|---------|---------|
| BFO (R3c) | Before | 5.5872 | 5.5872 | 13.8907 | 90 | 90 | 120 |
|  | After | 5.5216 | 5.5216 | 13.7287 | 90 | 90 | 120 |
| BFO (P1) | Before | 3.93100 | 3.93100 | 3.9310 | 90.3200 | 90.2400 | 89.9800 |
|  | After | 3.93104 | 3.93104 | 3.93104 | 90.3194 | 90.2443 | 89.9762 |

## Appendix D: DFT calculation of spontaneous polarisation

Spontaneous polarization stands as a major ferroelectric property of BFO. Hence, it is imperative to examine the spontaneous polarization of this material. In this study, we have computed the spontaneous polarization by considering the combined contributions of electronic and ionic polarizations. The DFT calculated polarisation and tilting angle in both the R3c and P1 BFO are represented in Figure 8(a-d). In the case of R3c-structured BFO, the spontaneous polarization arises from the displacement of Bi, Fe, and O atoms along the [111] direction [65]. Conversely, P1 BFO exhibits polarization along the [100] direction. This spontaneous polarization can be attributed to the electronic hybridization between the 6s electronic states of Bi, and the 2s and 2p states of O. This hybridization results in a stereo chemically active polarizable lone pair of Bi atoms, giving rise to the observed spontaneous polarization [66, 67]. The Quantum Espresso software utilizing the PBE sol pseudopotential was employed to compute the total spontaneous polarization. The calculated spontaneous polarization values are 72 µC/cm² and 102 µC/cm² for R3c and P1 BFO, respectively. The spontaneous polarization is shown schematically in Figure 8(a) and (b) for R3c and P1 respectively. Our computed polarization values of R3c BFO fall within the range of some previously reported theoretical results of R3c BFO, 87.8 µC/cm² [68], 90.9 µC/cm² [69], and 57.34 µC/cm² [70]. Interestingly, our calculations reveal much higher spontaneous polarization in the triclinic BFO as compared to the conventional rhombohedral BFO. Fig. 8 (c) and (d) shows the tilting angle ω with respect to the directions parallel to a and b crystallographic axis of the FeO₆ octahedra and the Fe-O-Fe bond angle θ [71]. The P1 BFO structure is more tilted than the R3c BFO.

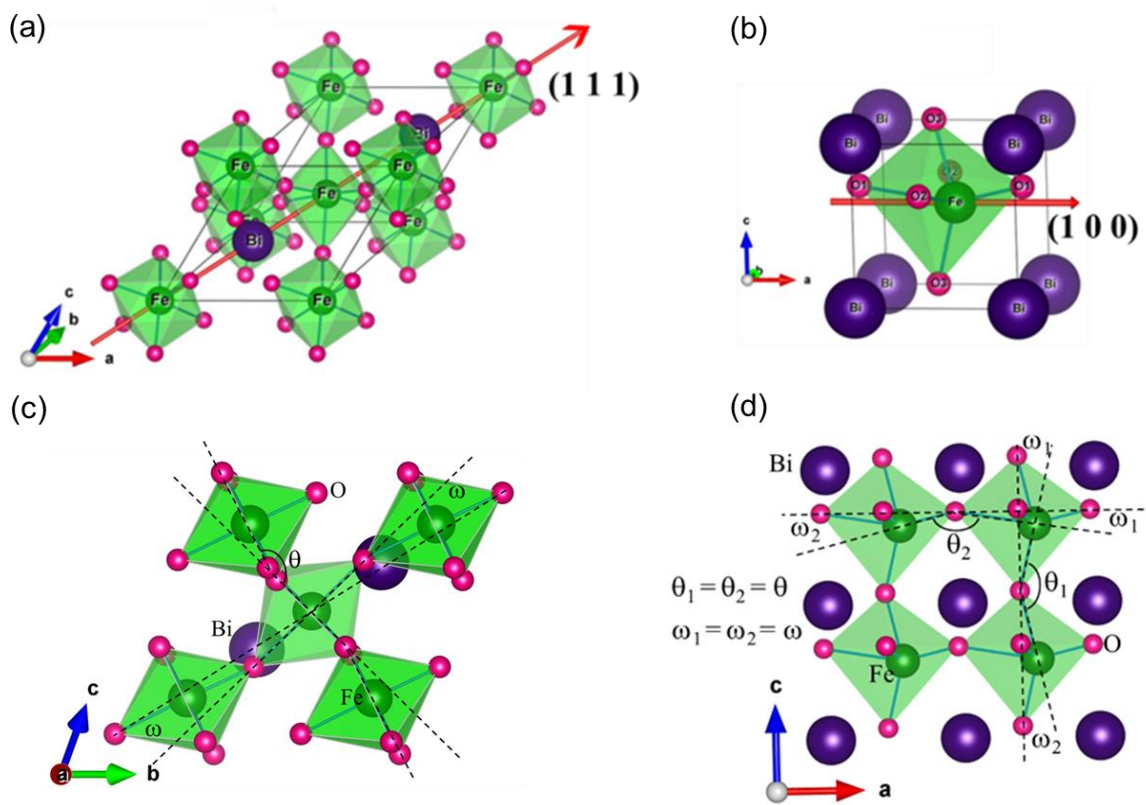

**Figure 8:** (a) The polarization direction of Bi- atoms and FeO₆ octahedra along (111) and, (b) (100) direction for R3c and P1 BFO structures respectively, (c) The octahedral tilting angle ω

and Fe-O-Fe bond angle θ for R3c BFO. (d) The octahedral tilting angle ω and Fe-O-Fe bond angle θ with respect to directions parallel to a- axis and c- axis for P1 BFO.

## References

[1] W.-R. Geng, Y.-L. Tang, Y.-L. Zhu, Y.-J. Wang, B. Wu, L.-X. Yang, Y.-P. Feng, M.-J. Zou, T.-T. Shi, Y. Cao, X.-L. Ma, Magneto–Electric–Optical Coupling in Multiferroic $BiFeO_3$-Based Films, Adv. Mater. **34**, 2106396 (2022).

[2] Y. L. Huang, D. Nikonov, C. Addiego, R. V. Chopdekar, B. Prasad, L. Zhang, J. Chatterjee, H. J. Liu, A. Farhan, Y. H. Chu, M. Yang, M. Ramesh, Z. Q. Qiu, B. D. Huey, C. C. Lin, T. Gosavi, J. Iniguez, J. Bokor, X. Pan, I. Young, L. W. Martin, and R. Ramesh, Manipulating magnetoelectric energy landscape in multiferroics, Nat. Commun. **11**, 2836 (2020).

[3] D. Sando, A. Barthelemy, and M. Bibes, $BiFeO_3$ epitaxial thin films and devices: past, present and future, J. Phys.: Condens. Matter **26**, 473201 (2014).

[4] G. Catalan, and J. F. Scott, Physics and applications of bismuth ferrite, Adv. Mater. **21**, 2463 (2009).

[5] D. Yi, P. Yu, Y. C. Chen, H. H. Lee, Q. He, Y. H. Chu, and R. Ramesh, Tailoring Magnetoelectric Coupling in $BiFeO_3/La_{0.7}Sr_{0.3}MnO_3$ Heterostructure through the Interface Engineering, *Adv. Mater.* **31**, 1806335 (2019).

[6] J. H. Lee, H. J. Choi, D. Lee, M. G. Kim, C. W. Bark, S. Ryu, M.-A. Oak, and H. M. Jang, Variations of ferroelectric of-centering distortion and 3d-4p orbital mixing in La-doped $BiFeO_3$ multiferroics, Phys. Rev. B **82**, 045113 (2010).

[7] L. Yin, and W. Mi, Progress in $BiFeO_3$-based heterostructures: materials, properties and applications, Nanoscale **12**, 477 (2020).

[8] J. Silva, A. Reyes, H. Esparza, H. Camacho, and L. Fuentes, $BiFeO_3$: A Review on Synthesis, Doping and Crystal Structure, Integr. Ferroelectr. **126**, 47 (2011).

[9] S. Manipatruni, D. E. Nikonov, C. C. Lin, T. A. Gosavi, H. Liu, B. Prasad, Y.-L. Huang, E. Bonturim, R. Ramesh, and I. A. Young, Scalable energy-efficient magnetoelectric spin–orbit logic, *Nature* **565**, 35 (2019).

[10] A. Haykal, J. Fischer, W. Akhtar, J. Y. Chauleau, D. Sando, A. Finco, F. Godel, Y. A. Birkholzer, C. Carretero, N. Jaouen, M. Bibes, M. Viret, S. Fusil, V. Jacques, and V. Garcia, Antiferromagnetic textures in $BiFeO_3$ controlled by strain and electric field, Nat. Commun. **11**, 1704 (2020).

[11] N. A. Spaldin, and R. Ramesh, Advances in magnetoelectric multiferroics. Nat. Mater. **18**, 203 (2019).

[12] C. R. Joshi, M. Acharya, G. J. Mankey, and A. Gupta, Effect of thickness and frequency of applied field on the switching dynamics of multiferroic bismuth ferrite thin films, Phys. Rev. Mater. 6, 054409 (2022).

[13] L. Yin, X. Wang, and W. Mi, Tunable Valley and Spin Polarizations in $BiXO_3/BiIrO_3$ (X = Fe, Mn) Ferroelectric Superlattices, ACS Appl. Mater. Interfaces **10**, 3822 (2018).

[14] R. Zhang, P. Hu, L. Bai, X. Xie, H. Dong, M. Wen, Z. Mu, X. Zhanga and F. Wu, New multiferroic $BiFeO_3$ with large polarization, Phys. Chem. Chem. Phys. **24**, 5939 (2022).

[15] S. Saha, R. P. Singh, Y. Liu, A. B. Swain, A. Kumar, V. Subramanian, A. Arockiarajan, G. Srinivasan, and R. Ranjan, Strain transfer in ferroelectric-ferrimagnetic magnetoelectric composite, Phys. Rev. B **103**, L140106 (2021).

[16] N. Wang, X. Luo, L. Han, Z. Zhang, R. Zhang, H. Olin and Y. Yang, Structure, Performance, and Application of $BiFeO_3$ Nanomaterials, Nano-Micro Lett. **12**, 81 (2020).

[17] R. I. Dass, J.Q. Yan, and J.B. Goodenough, Oxygen stoichiometry, ferromagnetism, and transport properties of $La_{2-x}NiMnO_{6+\delta}$, Phys. Rev. B **68**, 064415 (2003).

[18] M. S. Sheikh, D. Ghosh, A. Dutta, S. Bhattacharyya, and T. P. Sinha, Lead free double perovskite oxides $Ln_2NiMnO_6$ (Ln = La, Eu, Dy, Lu), a new promising material for photovoltaic application, Mater. Sci. Eng.: B **226**, 10 (2017).

[19] M. S. Sheikh, D. Ghosh, T. K. Bhowmik, A. Dutta, S. Bhattacharyya, and T. P. Sinha. When multiferroics become photoelectrochemical catalysts: A case study with $BiFeO_3/La_2NiMnO_6$, Mater. Chem. Phys. **244**, 122685 (2020).

[20] Z. Chen, S. Prosandeev, Z. L. Luo, W. Ren, Y. Qi, C. W. Huang, L. You, C. Gao, I. A. Kornev, T. Wu, J. Wang, P. Yang, T. Sritharan, L. Bellaiche, and L. Chen, Coexistence of ferroelectric triclinic phases in highly strained $BiFeO_3$ films, Phys. Rev. B **84**, 094116 (2011).

[21] M. R. Walden, C. V. Ciobanu, and G. L. Brennecka, Density-functional theory calculation of magnetic properties of $BiFeO_3$ and $BiCrO_3$ under epitaxial strain, J. Appl. Phys. **130**, 104102 (2021).

[22] B. Xu, S. Meyer, M. J. Verstraete, L. Bellaiche, and B. Dupé, First-principles study of spin spirals in the multiferroic $BiFeO_3$ Phys. Rev. B **103**, 214423 (2021).

[23] F. Sun, D. Chen, X. Gao, and J.-M. Liu, Emergent strain engineering of multiferroic $BiFeO_3$ thin films, J. Materiomics. **7**, 281 (2021).

[24] P. Blaha, K. Schwarz, F. Tran, R. Laskowski, G.K.H. Madsen, L.D. Marks, Wien2k, An apw+lo program for calculating the properties of solids, J. Chem. Phys. **152**, 074101, (2020).

[25] P. Blaha, K. Schwarz, P. Sorantin, S. Trickey, Full-potential, linearized augmented plane wave programs for crystalline systems, Comput. Phys. Commun. **59**, 399 (1990).

[26] S. Yahyaoui, S. Kallel, H. T. Diep, Magnetic properties of perovskites $La_{0.7}Sr_{0.3}Mn^{3+}_{0.7}Mn^{4+}_{0.3-x}Ti_xO_3$: Monte Carlo simulation versus experiments, J. Magn. Magn. Mater. **416**, 441 (2016).

[27] Md. S. Sheikh, A. Dutta, T.K. Bhowmik, S.K. Ghosh, S.K. Rout, T.P. Sinha, Synthesis, structural and photo physical properties of perovskite oxide $(KNbO_3)1-X+(La_2NiMnO_6)X$ for photovoltaic application, 35th European Photovoltaic Solar Energy Conference and Exhibition, Brussels, Belgium, 2018, pp. 143–148. https://doi.org/10.4229/35thEUPVSEC20182018-1CV.4.36.

[28] M.B. Mohamed, H. Wang, H. Fuess, Dielectric relaxation and magnetic properties of Cr doped $GaFeO_3$, J. Phys. D. Appl. Phys. **43**, 455409 (2010).

[29] A. K. Jonscher, Dielectric Relaxation in Solids, Chelsea Dielectrics Press, London, 1983.

[30] E. F. Hairetdinov, N. F. Uvarov, H. K. Patel, and S. W. Martin, Estimation of the free-charge carrier concentration in fast-ion conducting $Na_2S-B_2S_3$ glasses from an analysis of the frequency-dependent conductivity, Phys. Rev. B **50**, 13259 (1994).

[31] G. W. Pabst, L. W. Martin, Y.-H. Chu, R. Ramesh, Leakage mechanisms in $BiFeO_3$ thin films, Appl. Phys. Lett. **90**, 072902 (2007).

[32] A. Radmilovic, T. J. Smart, Y. Ping, K.-S. Choi, Combined Experimental and Theoretical Investigations of n-Type $BiFeO_3$ for Use as a Photoanode in a Photoelectrochemical Cell, Chem. Mater. **32**, 3262 (2020).

[33] G. Geneste, C. Paillard, B Dkhil, Polarons, vacancies, vacancy associations, and defect states in multiferroic $BiFeO_3$, Phys. Rev. B **99**, 024104 (2019).

[34] M. S. Sheikh, A. P. Sakhya, A. Dutta, and T. P. Sinha, Light induced charge transport in $La_2NiMnO_6$ based Schottky diode, J. Alloys Compd. **727**, 238 (2017).

[35] N. S. Rogado, J. Li, A. W. Sleight, and M. A. Subramanian, Magnetocapacitance and Magnetoresistance Near Room Temperature in a Ferromagnetic Semiconductor: $La_2NiMnO_6$, Adv. Mater. **17**, 2225 (2005).

[36] S. Zhao, L. Shia, S. Zhou, J. Zhao, H. Yang, and Y. Guo, Size-dependent magnetic properties and Raman spectra of $La_2NiMnO_6$ nanoparticles, J. Appl. Phys. **106**, 123901 (2009).

[37] N. Gao, C. Quan, Y. Ma, Y. Han, Z. Wu, W. Mao, J. Zhang, J. Yang, X. Li, W. Huang, Experimental and first principles investigation of the multiferroics $BiFeO_3$ and $Bi_{0.9}Ca_{0.1}FeO_3$: Structure, electronic, optical and magnetic properties, Physica B: Condens. Mat. **481**, 45 (2016).

[38] C. Ederer, and N. A. Spaldin, Weak ferromagnetism and magnetoelectric coupling in bismuth ferrite, Phys. Rev. B **71**, 060401(R) (2005).

[39] M. Rangi, A. Agarwal, S. Sanghi, R. Singh, S.S. Meena, and A. Das, Crystal structure and magnetic properties of $Bi_{0.8}A_{0.2}FeO_3$ (A = La, Ca, Sr, Ba) multiferroics using neutron diffraction and Mossbauer spectroscopy, AIP. Adv. **4**, 87121 (2014).

[40] P. K. Gupta, S. Ghosh, S. Kumar, A. Pal, P. Singh, M. Alam, A. Singh, S. Roy, R. Singh, B. P. Singh, N. N. Kumar, E. F. Schwier, M. Sawada; T. Matsumura, K. Shimada, H.-J. Lin, Y.-Y. Chin, A. K. Ghosh, S. Chatterjee, Room temperature exchange bias in antiferromagnetic composite $BiFeO_3$-$TbMnO_3$, J. Appl. Phys. **126**, 243903 (2019).

[41] M. A. Matin, M. N. Hossain, M. A. Hakim, M. F. Islam, Effects of Gd and Cr co-doping on structural and magnetic properties of $BiFeO_3$ nanoparticles, Mater. Res. Express **6**, 055038 (2019).

[42] Huimin Xian, Lingyun Tang, Zhongquan Mao, Jiang Zhang, Xi Chen, Bounded Magnetic Polarons Induced Enhanced Magnetism in Ca-doped $BiFeO_3$, Solid State Commun. **287**, 54 (2019).

[43] D. V. Karpinsky, O. M. Fesenko, M. V. Silibin, S. V. Dubkov, M. Chaika, A. Yaremkevich, A. Lukowiak, Y. Gerasymchuk, W. Stręk, A. Pakalniškis, R. Skaudzius, A. Kareiva, Y. M. Fomichov, V. V. Shvartsman, S. V. Kalinin, N. V. Morozovsky, A. N. Morozovska, Ferromagnetic-like behavior of $Bi_{0.9}La_{0.1}FeO_3$–KBr nanocomposites, Sci. Rep. **9**, 10417 (2019).

[44] S. Rizwan, M. Umar, Z. U. D. Babar, S. U. Awan, M. A. Rehman, Selenium-enriched flower-like of bismuth ferrite nanosheets assembly with associated magnetic properties, AIP Adv. **9**, 055025 (2019).

[45] R. Singh, P. K. Gupta, S. Kumar, A. G. Joshi, A. K. Ghosh, S. Patil, and S. Chatterjee, Enhancement in electrical and magnetic properties with Ti-doping in $Bi_{0.5}La_{0.5}Fe_{0.5}Mn_{0.5}O_3$, J. App. Phys. **121**, 154101 (2017).

[46] A.S. Erchidi Elyacoubi, R. Masrour, A. Jabar, Magnetocaloric effect and magnetic properties in $SmFe_{1-x}MnxO_3$ perovskite: Monte Carlo simulations, Solid State Commun. **271**, 39-43, (2018).

[47] Poorva Sharma, R. Masrour, A. Jabar, Jiyu Fan, Ashwini Kumar, Langsheng Ling, Chunlan Ma, Caixia Wang, Hao Yang, Structural and magnetocaloric properties of rare-earth orthoferrite perovskite: $TmFeO_3$, Chem. Phys. Lett. **740**, 137057 (2020).

[48] T. K. Bhowmik, and T. P. Sinha, Al-dependent electronic and magnetic properties of $YCrO_3$ with magnetocaloric application: An ab-initio and Monte Carlo approach, Physica B: Condensed Matter **606**, 412659 (2021).

[49] G. L. Yuan, Siu Wing, J. M. Liu, and Z. G. Liu, Structural transformation and ferroelectromagnetic behavior in single-phase $Bi_{1-x}Nd_xFeO_3$ multiferroic ceramics, Appl. Phys. Lett. **89**, 052905 (2006).

[50] S.-W. Cheong, M. Mostovoy, Multiferroics: a magnetic twist for ferroelectricity, Nat. Mater. **6**, 13 (2007).

[51] J. C. Burtfoot. Ferroelectrics: An Introduction to the Physical Principles, Vân Nostrand-Reinbold, London (1967).

[52] S. Sharma, M. Kumar, Band gap tuning and optical properties of $BiFeO_3$ nanoparticles, Mater. Today: Proc. **28**, 168 (2020).

[53] T. Choi, S. Lee, Y. J. Choi, V. Kiryukhin, and W. Cheong, Switchable Ferroelectric Diode and Photovoltaic Effect in $BiFeO_3$, Science, **324**, 5923, 63-66 (2009).

[54] Ze Li, Yu Zhao, Wei-Li Li, Ruixuan Song, Wenyue Zhao, Zhao Wang, Yazhou Peng, and Wei-Dong Fei, J. Phys. Chem. C, **125 (17)**, 9411-9418, (2021).

[55] Sven Rühle, Tabulated values of the Shockley–Queisser limit for single junction solar cells, Solar Energy, **130**, 139-147, (2016).

[56] I. Sosnowska, T. P. Neumaier, and E. Steichele, Spiral magnetic ordering in bismuth ferrite, J. Phys. C: Solid State Phys. **15**, 4835 (1982).

[57] P. Hermet, M. Goffinet, J. Kreisel, and Ph. Ghosez, Raman and infrared spectra of multiferroic bismuth ferrite from first principles, Phys. Rev. B **75**, 220102(R) (2007).

[58] S. V. Streltsov, Magnetic moment suppression in $Ba_3CoRu_2O_9$: Hybridization effect, Phys. Rev. B **88**, 024429 (2013).

[59] J. T. Zhang, X. M. Lu, J. Zhou, H. Sun, J. Su, C. C. Ju, F. Z. Huang, and J. S. Zhu, Origin of magnetic anisotropy and spiral spin order in multiferroic $BiFeO_3$, Appl. Phys. Lett. 100, 242413 (2012).

[60] P. Fischer, M. PoIomska, I. Sosnowska, M. Szymanski, Temperature dependence of the crystal and magnetic structures of $BiFeO_3$, J. Phys. C: Solid St. Phys. **13**, 1931 (1980).

[61] R. Masrour, and J. Abderrahim, Magnetocaloric and magnetic properties of $La_2NiMnO_6$ double perovskite, Chin. Phys. B, **25**, 087502 (2016).

[62] D. P. Landau, K. Binder, A Guide to Monte Carlo Simulations in Statistical Physics, Third Edition, Cambridge University Press, New York, 2009.

[63] M. E. J. Newman, G. T. Barkema, Monte Carlo Methods in Statistical Physics, Oxford University Press Inc., New York, 1999.

[64] P. Giannozzi et al., QUANTUM ESPRESSO: a modular and open-source software project for quantum simulations of materials, J. Phys.: Condens. Matter**. 21,** 395502 (2009).

[65] D. Lebeugle, D. Colson, A. Forget, M. Viret, Very large spontaneous electric polarization in $BiFeO_3$ single crystals at room temperature and its evolution under cycling fields, Appl. Phys. Lett. 91, 022907 (2007).

[66] P. Ravindran, R. Vidya, A. Kjekshus, H. Fjellvåg, and O. Eriksson, Theoretical investigation of magnetoelectric behavior in BiFeO$_3$, Phys. Rev. B **74**, 224412 (2006).

[67] M. Tokunaga, M. Akaki, T. Ito, S. Miyahara, A. Miyake, H. Kuwahara, N. Furukawa, Magnetic control of transverse electric polarization in BiFeO$_3$. Nat Commun. **6**, 5878 (2015).

[68] A. Stroppa, S. Picozzi, Hybrid functional study of proper and improper multiferroics, Phys. Chem. Chem. Phys. **12**, 5405 (2010).

[69] Pugaczowa-Michalska, M., Kaczkowski, J. First-principles study of structural, electronic, and ferroelectric properties of rare-earth-doped BiFeO$_3$, J. Mater. Sci. **50**, 6227 (2015).

[70] L. H. da S. Lacerda, R. A. P. Ribeiro, S. R. de Lazaro, Magnetic, electronic, ferroelectric, structural and topological analysis of AlFeO$_3$, FeAlO$_3$, FeVO$_3$, BiFeO$_3$ and PbFeO$_3$ materials: Theoretical evidences of magnetoelectric coupling, J. Magn. Magn. Mater. **480**, 199 (2019).

[71] H. H'Mõk, E. M. Aguilar, J. A. García, J. R. Ariño, L. Mestres, P. Alemany, D.H. Galván, J.M. S. Beltrones, O. R. Herrera, Theoretical justification of stable ferromagnetism in ferroelectric BiFeO$_3$ by first-principles, Comput. Mater. Sci. **164**, 66 (2019).